\documentstyle[12pt]{article}

\catcode`\@=11
\long\def\@makefntext#1{
\protect\noindent \hbox to 3.2pt {\hskip-.9pt
$^{{\ninerm\@thefnmark}}$\hfil}#1\hfill}                

 \def\@makefnmark{\hbox to 0pt{$^{\@thefnmark}$\hss}}  

\def\ps@myheadings{\let\@mkboth\@gobbletwo
\def\@oddhead{\hbox{}
\rightmark\hfil\ninerm\thepage}
\def\@oddfoot{}\def\@evenhead{\ninerm\thepage\hfil
\leftmark\hbox{}}\def\@evenfoot{}
\def\sectionmark##1{}\def\subsectionmark##1{}}

\newcounter{sectionc}\newcounter{subsectionc}\newcounter{subsubsectionc}
\renewcommand{\section}[1] {\vspace{0.6cm}\addtocounter{sectionc}{1}
\setcounter{subsectionc}{0}\setcounter{subsubsectionc}{0}\noindent
	{\bf\thesectionc. #1}\par\vspace{0.4cm}}
\renewcommand{\subsection}[1] {\vspace{0.6cm}\addtocounter{subsectionc}{1}
	\setcounter{subsubsectionc}{0}\noindent
	{\it\thesectionc.\thesubsectionc. #1}\par\vspace{0.4cm}}
\renewcommand{\subsubsection}[1]
{\vspace{0.6cm}\addtocounter{subsubsectionc}{1}
	\noindent {\rm\thesectionc.\thesubsectionc.\thesubsubsectionc.
	#1}\par\vspace{0.4cm}}

\newcounter{appendixc}
\newcounter{subappendixc}[appendixc]
\newcounter{subsubappendixc}[subappendixc]

\renewcommand{\appendix}[1] {\vspace{0.6cm}
	\refstepcounter{appendixc}
	\setcounter{figure}{0}
	\setcounter{table}{0}
	\setcounter{equation}{0}
	\renewcommand{\thefigure}{\Alph{appendixc}.\arabic{figure}}
	\renewcommand{\thetable}{\Alph{appendixc}.\arabic{table}}
	\renewcommand{\theappendixc}{\Alph{appendixc}}
	\renewcommand{\theequation}{\Alph{appendixc}.\arabic{equation}}
	\noindent{\bf Appendix \theappendixc #1}\par\vspace{0.4cm}}

\def\abstracts#1{{
	\centering{\begin{minipage}{30pc}\tenrm\baselineskip=12pt\noindent
	\centerline{\tenrm ABSTRACT}\vspace{0.3cm}
	\parindent=0pt #1
	\end{minipage}}\par}}

\renewenvironment{thebibliography}[1]
	{\begin{list}{\arabic{enumi}.}
	{\usecounter{enumi}\setlength{\parsep}{0pt}
\setlength{\leftmargin 1.25cm}{\rightmargin 0pt}
	 \setlength{\itemsep}{0pt} \settowidth
	{\labelwidth}{#1.}\sloppy}}{\end{list}}

\newcounter{itemlistc}
\newcounter{romanlistc}
\newcounter{alphlistc}
\newcounter{arabiclistc}

\newcommand{\fcaption}[1]{
	\refstepcounter{figure}
	\setbox\@tempboxa = \hbox{\tenrm Fig.~\thefigure. #1}
	\ifdim \wd\@tempboxa > 6in
	   {\begin{center}
	\parbox{6in}{\tenrm\baselineskip=12pt Fig.~\thefigure. #1}
	    \end{center}}
	\else
	     {\begin{center}
	     {\tenrm Fig.~\thefigure. #1}
	      \end{center}}
	\fi}

\newcommand{\tcaption}[1]{
	\refstepcounter{table}
	\setbox\@tempboxa = \hbox{\tenrm Table~\thetable. #1}
	\ifdim \wd\@tempboxa > 6in
	   {\begin{center}
	\parbox{6in}{\tenrm\baselineskip=12pt Table~\thetable. #1}
	    \end{center}}
	\else
	     {\begin{center}
	     {\tenrm Table~\thetable. #1}
	      \end{center}}
	\fi}

\def\@citex[#1]#2{\if@filesw\immediate\write\@auxout
	{\string\citation{#2}}\fi
\def\@citea{}\@cite{\@for\@citeb:=#2\do
	{\@citea\def\@citea{,}\@ifundefined
	{b@\@citeb}{{\bf ?}\@warning
	{Citation `\@citeb' on page \thepage \space undefined}}
	{\csname b@\@citeb\endcsname}}}{#1}}

\newif\if@cghi
\def\cite{\@cghitrue\@ifnextchar [{\@tempswatrue
	\@citex}{\@tempswafalse\@citex[]}}
\def\citelow{\@cghifalse\@ifnextchar [{\@tempswatrue
	\@citex}{\@tempswafalse\@citex[]}}
\def\@cite#1#2{{$\null^{#1}$\if@tempswa\typeout
	{IJCGA warning: optional citation argument
	ignored: `#2'} \fi}}

\def\fnt#1#2{\footnotetext{\kern-.3em
	{$^{\mbox{\sevenrm #1}}$}{#2}}}

 1
 1
 1

\font\tenrm=cmr10
\font\tenit=cmti10

\font\ninerm=cmr9

\begin{document}

\begin{flushright} UCRHEP-T138\\November 1994\
\end{flushright}
\vspace{0.8cm}

\centerline{\bf LEFT-RIGHT GAUGE SYMMETRY}
\baselineskip=16pt
\centerline{\bf AT THE TEV ENERGY SCALE\footnote{To appear in the
Proceedings of the 7th Adriatic Meeting on Particle Physics, Brijuni,
Croatia (September 1994)}}
\vspace{0.8cm}
\centerline{\tenrm ERNEST MA}
\baselineskip=13pt
\centerline{\tenit Department of Physics, University of California,}
\baselineskip=12pt
\centerline{\tenit Riverside, California 92521, USA}
\baselineskip=13pt
\vspace{0.9cm}
\abstracts{Two first examples beyond the standard model are given which
exhibit left-right symmetry ($g_L = g_R$) and supersymmetry at a few TeV,
together with gauge-coupling unification at around $10^{16}$ GeV.}
\vspace{0.8cm}
\rm\baselineskip=14pt
\section{Introduction}
What lies beyond the standard model at or below the TeV energy scale?
One very well-motivated possibility is supersymmetry.  In particular, the
minimal supersymmetric standard model (MSSM) is being studied by very
many people.  Another possibility is left-right gauge symmetry, but there
are a lot fewer advocates here and for good reason, as I will explain in
this talk.  I will also discuss how these problems may be overcome,
assuming both supersymmetry and left-right gauge symmetry at the TeV
energy scale.
\vspace{0.2cm}

There are two problems with the conventional left-right gauge model at the
TeV energy scale with or without supersymmetry.  One is the
unavoidable occurrence of flavor-changing neutral currents (FCNC) at tree
level.  The other is the lack of gauge-coupling unification which is known
to be well satisfied by the MSSM.\cite{1}  In this talk, I will offer two
new models.\cite{2,3}  Both allow the gauge couplings to be unified at around
$10^{16}$ GeV.  The second has the added virtue of being free of FCNC
at tree level.  Hence left-right gauge symmetry at a few TeV should be
considered a much more
attractive possibility than was previously recognized.

\section{Origin of FCNC in Left-Right Models}
Consider the gauge symmetry $\rm SU(3)_C \times SU(2)_L \times SU(2)_R
\times U(1)_{B-L}$ which breaks down to the standard $\rm SU(3)_C \times
SU(2)_L \times U(1)_Y$ at $M_R \sim$ few TeV with particle content
given by
\begin{eqnarray}
Q \equiv \left( \! \begin{array} {c} u \\ d \end{array} \! \right)_L \sim
(3,2,1,1/6), &~& Q^c \equiv \left( \! \begin{array} {c} d^c \\ u^c \end{array}
\! \right)_L \sim (\overline 3,1,2,-1/6), \\ L \equiv \left( \! \begin{array}
{c} \nu \\ l \end{array} \! \right)_L \sim (1,2,1,-1/2), &~& L^c \equiv
\left( \! \begin{array} {c} l^c \\ \nu^c \end{array} \! \right)_L \sim
(1,1,2,1/2).
\end{eqnarray}
Note that each generation of quarks and leptons ({\it i.e.} $Q + Q^c + L
+ L^c$) fits naturally into a {\bf 16} representation of SO(10).  In order
for the quarks and leptons to obtain nonzero masses, a scalar bidoublet
\begin{equation}
\eta \equiv \left[ \begin{array} {c@{\quad}c} \eta_1^0 & \eta_2^+ \\
\eta_1^- & \eta_2^0 \end{array} \right] \sim (1,2,2,0)
\end{equation}
is required.  Consider the interaction of $\eta$ with the quarks:
\begin{equation}
Q Q^c \eta = d d^c \eta_1^0 - u d^c \eta_1^- + u u^c \eta_2^0 -
d u^c \eta_2^+.
\end{equation}
If there is just one $\eta$, then the mass matrices for the $u$ and $d$
quarks are related by
\begin{equation}
{\cal M}_d \langle \eta_1^0 \rangle^{-1} = {\cal M}_u \langle \eta_2^0
\rangle^{-1},
\end{equation}
which means that there can be no mixing among generations and the ratio
$m_u/m_d$ is the same for each generation.  This is certainly not
realistic and two $\eta$'s will be required.
\begin{eqnarray}
{\cal M}_d &=& f \langle \eta_1^0 \rangle + f' \langle \eta_1'^0 \rangle, \\
{\cal M}_u &=& f \langle \eta_2^0 \rangle + f' \langle \eta_2'^0 \rangle.
\end{eqnarray}
As a result, the diagonalizations of ${\cal M}_u$ and ${\cal M}_d$ do not also
diagonalize the respective Yukawa couplings, hence FCNC are unavoidable.
To suppress these contributions to processes such as $K^0 - \overline {K^0}$
mixing, the fine tuning of couplings is required if $M_R \sim$ few TeV.
In the nonsupersymmetric case, $\eta'$ can be simply taken to be
$\sigma_2 \eta^* \sigma_2$, but that will not alleviate the FCNC problem.
Similarly, if the $f'$ terms were radiative corrections from, say, soft
supersymmetry breaking, FCNC would still be present.

\section{Evolution of Gauge Couplings}
Consider now the evolution of the gauge couplings to one-loop order.
\begin{equation}
\alpha_i^{-1} (M_U) = \alpha_i^{-1} (M_R) - {b_i \over {2\pi}} \ln {M_U
\over M_R},
\end{equation}
where $\alpha_i \equiv g_i^2/4 \pi$ and $b_i$ are constants determined by
the particle content contributing to $\alpha_i$.  Using the standard model
to evolve $\alpha_i$ from their experimentally determined values at $M_Z$ to
$M_R \sim$ few TeV and requiring that they converge to a single value at
around $10^{16}$ GeV, the constraints
\begin{equation}
b_2 - b_3 \sim 4, ~~~ b_1 - b_2 \sim 14,
\end{equation}
are obtained.  It is easily seen that these constraints are not
satisfied by the conventional left-right gauge model with or without
supersymmetry.  Note that $b_2 - b_3 = 4$ in the MSSM, corresponding
to two $\rm SU(2)_L$ doublets, whereas in the supersymmetric left-right
model with two bidoublets (four $\rm SU(2)_L$ doublets), $b_2 - b_3 = 5$.

\section{First Example with Unification}
Suppose the FCNC problem is disregarded, then the conventional left-right
model with particle assignments given by Eqs. (1) and (2) can be made to
have gauge-coupling unification if new particles are added at the TeV
energy scale.\cite{2}  Supersymmetry is also assumed so that $M_R$
and $M_U$ can be separated naturally.  Now
\begin{eqnarray}
b_S &=& - 9 + 2(3) + n_D = -1, \\ b_{LR} &=& - 6 + 2(3) + n_{22} + n_H
= 3, \\ (3/2) b_X &=& 2(3) + 3n_H + n_D + 3n_E = 17,
\end{eqnarray}
and the constraints of Eq. (9) are satisfied.  The gauge couplings do
meet at one point as shown in Fig. 1, based on a full two-loop
numerical analysis.

In this model $n_{22} = 2$ is the number of bidoublets, $n_H = 1$ is
the number of an anomaly-free set of Higgs doublets needed to break
the $\rm SU(2)_R$ symmetry independent of $\rm SU(2)_L$:
\begin{eqnarray}
\Phi_L \sim (1,2,1,-1/2), &~& \Phi_R \sim (1,1,2,1/2), \\ \Phi_L^c \sim
(1,2,1,1/2), &~& \Phi_R^c \sim (1,1,2,-1/2),
\end{eqnarray}
$n_D = 2$ is the number of exotic singlet quarks of charge $-1/3$:
\begin{equation}
D \sim (3,1,1,-1/3), ~~~ D^c \sim (\overline 3, 1,1,1/3),
\end{equation}
and $n_E = 2$ is the number of exotic singlet leptons of charge $-1$:
\begin{equation}
E \sim (1,1,1,-1), ~~~ E^c \sim (1,1,1,1).
\end{equation}
Note that $n_{22} = 2$ and $n_H = 1$ are required for fermion masses and
$\rm SU(2)_R$ breaking respectively.  To obtain $b_{LR} - b_S = 4$,
$n_D = 2$ is then assumed.  At this stage, $(3/2)b_X - b_{LR} = 8$.
To increase that to 14, $n_E = 2$ is just right.  This should not be
considered fine tuning because the contribution of each new set of particles
comes in large chunks, 3 in the case of the $E$'s for example; so if 6
did not happen to be the desired number, it would not have been possible
to achieve unification with the addition of new particles this way.

\section{Left-Right Model without FCNC}
Consider the E$_6$ superstring-inspired left-right model proposed some
years ago.\cite{4,5}  In the fundamental {\bf 27} representation of E$_6$,
there is an additional quark singlet of charge $-1/3$.  An alternative
to the conventional left-right assignment is then possible:
\begin{eqnarray}
Q \equiv \left( \! \begin{array} {c} u \\ d \end{array} \! \right)_L
\sim (3,2,1,1/6), &~& d_L^c \sim (\overline 3, 1,1,1/3), \\ Q^c \equiv
\left( \! \begin{array} {c} h^c \\ u^c \end{array} \! \right)_L \sim
(\overline 3, 1,2,-1/6), &~& h_L^c \sim (3,1,1,-1/3),
\end{eqnarray}
where the switch $h^c$ for $d^c$ has been made. The doublets
$\Phi_{L,R}$ and the bidoublet $\eta$ are also in the {\bf 27}.
Hence the following terms are allowed:
\begin{eqnarray}
Q Q^c \eta &=& d h^c \eta_1^0 - u h^c \eta_1^- + u u^c \eta_2^0 -
d u^c \eta_2^+, \\ Q d^c \Phi_L &=& d d^c \phi_L^0 - u d^c \phi_L^-, \\
h Q^c \Phi_R &=& h h^c \phi_R^0 - h u^c \phi_R^+.
\end{eqnarray}
As a result,
\begin{equation}
{\cal M}_u \propto \langle \eta_2^0 \rangle, ~~~ {\cal M}_d \propto
\langle \phi_L^0 \rangle, ~~~ {\cal M}_h \propto \langle \phi_R^0 \rangle.
\end{equation}
Since each quark type has its own source of mass generation, FCNC are now
guaranteed to be absent at tree level.  This is the only example of a
left-right model without FCNC.

\section{Extended Definition of Lepton Number}
Since the (1,2,1,$-1/2$) component of the {\bf 27} is now identified
as the Higgs superfield $\Phi_L$, where are the leptons of this model?
One lepton doublet is in fact contained in the bidoublet,
{\it i.e.}~$(\nu,l)_L$ should be
identified with the spinor components of $(\eta_1^0,
\eta_1^-)$, and one lepton singlet $l^c_L$ with that of $\phi_R^+$.  Since
\begin{equation}
\Phi_L \Phi_R \eta = \phi_L^- \phi_R^+ \eta_1^0 - \phi_L^0 \phi_R^+ \eta_1^-
+ \phi_L^0 \phi_R^0 \eta_2^0 - \phi_L^- \phi_R^0 \eta_2^+,
\end{equation}
the lepton $l$ gets a mass from $\langle \phi_L^0 \rangle$.  Furthermore,
from Eq. (19), it is seen that the exotic quark $h$ must have lepton
number $L = 1$ and since $u^c$ and $h^c$ are linked by $\rm SU(2)_R$,
the $W_R^-$ gauge boson must also have $L = 1$.  This extended definition
of lepton number is consistent with all the interactions of this model
and is conserved.
\vspace{0.2cm}

The production of $W_R$ in this model\cite{6,7} is very different from
that of the conventional left-right model.  Because of lepton-number
conservation, the best scenario is to have $u + g \rightarrow h + W_R^+$,
where $g$ is a gluon.  The decay of $W_R$ must end up with a lepton as
well as a particle with odd $R$ parity.  Note also that $W_L - W_R$ mixing
is strictly forbidden and $W_R$ does not contribute to $\Delta m_K$ or
$\mu$ decay.
\vspace{0.2cm}

Since the absence of FCNC allows only one bidoublet, only one lepton
generation is accounted for in the above.  Let it be the $\tau$ lepton.
The $e$ and $\mu$ generations are then accommodated in the $\Phi_{L,R}$
components of the other two {\bf 27}'s, but they must not couple to
$Q d^c$ or $h Q^c$.  This can be accomplished by extending the
discrete symmetry necessary for maintaining the conservation of lepton
number as defined above.\cite{3}

\section{Precision Measurements at the Z}
Because of the Higgs structure of this model, there is in general some
$Z- Z'$ mixing which depends on the ratio of the $W_L$ to $W_R$ masses.
Let $\langle \eta_2^0 \rangle = v$, $\langle \phi_{L,R}^0 \rangle =
v_{L,R}$, $r = v^2/(v^2 + v_L^2)$, $x = \sin^2 \theta_W$, then
\begin{equation}
M^2_{W_{L,R}} = {1 \over 2} g^2 (v^2 + v^2_{L,R}),
\end{equation}
and
\begin{equation}
M_Z^2 \simeq {M^2_{W_L} \over {1-x}} \left[ 1 - \left( r - {x \over {1-x}}
\right)^2 \xi \right], ~~~ M^2_{Z'} \simeq {{1-x} \over {1-2x}} M^2_{W_R},
\end{equation}
where $\xi = M^2_{W_L}/M^2_{W_R}$.  Deviations from the standard model
can now be expressed in terms of the three oblique parameters
$\epsilon_{1,2,3}$ or $S,T,U$.  Using the precision experimental inputs
$\alpha, G_F, M_Z$, and the $Z \rightarrow e^-e^+, \mu^- \mu^+$ (but not
$\tau^- \tau^+$) rates and forward-backward asymmetries, they are given by
\begin{eqnarray}
\epsilon_1 = \alpha T &=& - \left( {{2-3x} \over {1-x}} -r \right) \left(
r - {x \over {1-x}} \right) \xi, \\ \epsilon_2 = - {{\alpha U} \over {4x}}
&=& - \left( r - {x \over {1-x}} \right) \xi, \\ \epsilon_3 = {{\alpha S}
\over {4x}} &=& - \left( {{1-2x} \over {2x}} \right) \left( r - {x \over
{1-x}} \right) \xi.
\end{eqnarray}
Note that the ratio $S/T$ must be positive and of order unity here.
Experimentally, $S,T,U$ are all consistent with being zero within about
1$\sigma$, but the central $S$ and $T$ values are $-0.42$
and $-0.35$ respectively.\cite{8}  These imply that $r \sim 0.8$ and
$\xi \sim 6 \times 10^{-3}$, hence the $W_R$ mass should be about 1 TeV
whcih is exactly consistent with this model's assumed $\rm SU(2)_R$
breaking scale.

In this model, the $\tau$ generation transforms differently under
$\rm SU(2)_R$, hence there is a predicted difference in the $\rho_l$ and
$\sin^2 \theta_l$ parameters governing $Z \rightarrow l^-l^+$ decay.
Specifically,
\begin{equation}
\rho_\tau - \rho_{e,\mu} = 2 \left( r - {x \over {1-x}} \right) \xi \sim
6 \times 10^{-3},
\end{equation}
compared with the experimental value of $0.0064 \pm 0.0048$, and
\begin{equation}
\sin^2 \theta_\tau - \sin^2 \theta_{e,\mu} = - x \left( r - {x \over {1-x}}
\right) \xi \sim - 7 \times 10^{-4},
\end{equation}
compared with the experimental value of $-0.0043 \pm 0.0022$.  The standard
model's prediction for either quantity is of course zero.

\section{Second Example with Unification}
Fig. 2 shows the two-loop evolution of gauge couplings
corresponding to the following situation.
Let the particle content of the proposed left-right model be restricted
to only components of the {\bf 27} and {\bf 27*} representations of E$_6$,
then unification is achieved with\cite{3}
\begin{eqnarray}
b_S &=& -9 + 2(3) + n_h = 0, \\ b_{LR} &=& -6 + 2(3) + n_{22} + n_\phi
= 4, \\ (3/2) b_X &=& 2(3) + n_h + 3n_\phi = 18,
\end{eqnarray}
where $n_h = 3$ and $n_{22} = 1$ are required as already discussed, and
$n_\phi = 3$ is the number of extra sets of $\Phi_L + \Phi_R + \Phi_L^c +
\Phi_R^c$.  Note that at least one such set is needed for $\rm SU(2)_R$
breaking and that the two constraints of Eq. (9) are simultaneously
satisfied with the one choice of $n_\phi = 3$.
\vspace{0.2cm}

To complete the model, six singlets $N \sim (1,1,1,0)$ are also assumed.
At the unification scale $M_U$, there are presumably six {\bf 27}'s and
three {\bf 27*}'s of E$_6$, which is then broken down to supersymmetric
$\rm SU(3)_C \times SU(2)_L \times SU(2)_R \times U(1)$ supplemented by
a discrete $Z_4 \times Z_2$ symmetry\cite{3}.  Of the three {\bf 27}'s
and three {\bf 27*}'s, only the combinations $\Phi_L + \Phi_R + \Phi_L^c
+ \Phi_R^c$ survive.  Of the other three {\bf 27}'s, only two bidoublets
do not survive.  At $M_R \sim$ few TeV, $\Phi_R$ and $\Phi_R^c$ break
$\rm SU(2)_R \times U(1)$ down to $\rm U(1)_Y$.  Supersymmetry is also
broken softly at $M_R$.  The surviving model at the electroweak energy
scale is the standard model with two Higgs doublets but not those of the
MSSM, as already explained in my first talk\cite{9} at this meeting.

\section{Lepton Masses}
The $\tau$ gets its mass from the $\Phi_L \Phi_R \eta$ term, but there can
be no such term for the $e$ and $\mu$.  Hence the latter two are massless
at tree level.  However, the soft supersymmetry-breaking term $\Phi_L
\Phi_R \tilde \eta$ (where $\tilde \eta = \sigma_2 \eta^* \sigma_2$ and
all three fields are scalars) is allowed, hence $m_e$ and $m_\mu$ are
generated radiatively from the mass of the U(1) gauge fermion.\cite{10}
The neutrinos obtain small seesaw masses from their couplings with the
three $N_L$'s which are assumed to have large Majorana masses.  The
$\nu_\tau N_L$ mass comes from the $\eta \eta N_L$ term, and the $\nu_e
N_L, \nu_\mu N_L$ masses come from the $\Phi_L \Phi_L^c N_L$ terms.

\section{Conclusion}
New physics in the framework of left-right gauge symmetry is possible at
the TeV energy scale even if grand unification is required.  Two examples
have been given, the second of which is particularly attractive: it is
free of FCNC at tree level and has negative contributions to the oblique
parameters $S$ and $T$ consistent with present experimental central values.

\section{Acknowledgements}
I thank Profs. D. Tadic and I. Picek and the other organizers of the 7th
Adriatic Meeting on Particle Physics for their great hospitality and a very
stimulating program.  This work was supported in part by the U. S.
Department of Energy under Grant No. DE-FG03-94ER40837.

\end{document}